\documentclass[preprint,showpacs,preprintnumbers,amsmath,amssymb]{revtex4-1}
\usepackage{latexsym}
\usepackage{epsfig}
\usepackage{float}
\usepackage{lipsum}
\usepackage{graphicx,times}
\usepackage{amssymb,amsmath}
\usepackage{color}
\usepackage{dcolumn}
\usepackage{epstopdf}
\usepackage{pgfplots}
\usepackage[colorlinks=true, urlcolor=blue, linkcolor=red, citecolor=blue]{hyperref}
\newcommand{\be}{\begin{equation}}
\newcommand{\ee}{\end{equation}}
\begin{document}
\title{Coupled-channels Faddeev AGS calculation of $K^{-}ppn$ and $K^{-}ppp$ quasi-bound states}

\author{S.~Marri}
\email{s.marri@ph.iut.ac.ir}
\affiliation{Department of Physics, Isfahan University of Technology, Isfahan 84156-83111, Iran}
\author{S. Z.~Kalantari}
\affiliation{Department of Physics, Isfahan University of Technology, Isfahan 84156-83111, Iran}
\date{\today}
\begin{abstract}
Using separable $\bar{K}N-\pi\Sigma$ potentials in the Faddeev equations, we calculated the 
binding energies and widths of the $K^{-}pp$, $K^{-}ppn$ and $K^{-}ppp$ quasi-bound states 
on the basis of three- and four-body Alt-Grassberger-Sandhas equations in the momentum 
representation. One- and two-pole version of $\bar{K}N-\pi\Sigma$ interaction are considered 
and the dependence of the resulting few-body energy on the two-body $\bar{K}N-\pi\Sigma$ 
potential was investigated. The $s$-wave [3+1] and [2+2] sub-amplitudes are obtained by using 
the Hilbert-Schmidt expansion procedure for the integral kernels. As a result, we found a 
four-body resonance of the $K^{-}ppn$ and $K^{-}ppp$ quasi-bound states with a binding energy 
in the range $B_{K^{-}ppn}\sim{55-70}$ and $B_{K^{-}ppp}\sim{90-100}$ MeV, respectively. The 
calculations yielded full width of $\Gamma_{K^{-}ppn}\sim{16-20}$ and $\Gamma_{K^{-}ppp}\sim{7-20}$ 
MeV.
\end{abstract}
\pacs{
}
\maketitle
%%%%%%%%%%%%%%%%%%%%%%%%%%%%%%%%%%%%%%%%%%%%%%%%%%%%%%%%%%%%%%%%%%%%%%%%%%%%%%%%%%%%%%%%%%%%%%%%%%%
\section{Introduction}
\label{intro}
%%%%%%%%%%%%%%%%%%%%%%%%%%%%%%%%%%%%%%%%%%%%%%%%%%%%%%%%%%%%%%%%%%%%%%%%%%%%%%%%%%%%%%%%%%%%%%%%%%%
The dynamics of antikaon interacting with nucleons and nuclei is one of the current challenging 
problems in strangeness nuclear physics. The $\bar{K}N$ interaction at low energy is strongly attractive and 
generates the $\Lambda$(1405) resonance (abbreviated as $\Lambda^*$) as a quasi-bound state embedded 
in the $\pi\Sigma$ continuum below the $\bar{K}N$ threshold. Thus, one expects unusual, interesting 
phenomena to be observed when the antikaon is injected or stopped in nuclei. Theoretical interests 
in $\bar{K}$-nuclear bound states were triggered by the works of Akaishi and Yamazaki (A-Y) looking 
for $\bar{K}$ bound states in several few-body systems~\cite{akaishi,yamazaki1,dote1,dote2}, which 
were predicted to be not only deeply bound but also unusually shrunk. In addition to the lightest possible 
antikaon-nucleus system, $K^{-}pp$, a series of proton-rich $K^{-}$ bound systems were predicted \cite{yamazaki1}, 
which can be called kaonic nuclear clusters (``KNC"). The proton and neutron distributions in KNC's were 
studied extensively using antisymmetrized molecular dynamics (AMD) method by Dote {\it et al.} \cite{dote1,dote2}. 
Subsequently, theoretical studies of KNC's, especially of $K^-pp$, were developed by using different 
models and methods to solve the three-body system \cite{shev1,shev2,ikeda1,ikeda2,dote3,dote4,ikeda3}. 
These calculations have shown essentially that the $K^-pp$ system is bound below the break-up 
threshold in agreement with A-Y's original prediction~\cite{yamazaki1}, though some differences between 
different predictions remain. Very recently, Maeda {\it et al.}~\cite{maeda} has carried out Faddeev and 
Faddeev-Yakubowsky calculations for the three and four body systems, $\bar{K}NN$, $\bar{K}NNN$, 
$\bar{K}\bar{K}N$ and $\bar{K}\bar{K}NN$, with varied elementary potentials, overviewing their binding 
energies, densities and shapes. 
 
It was found and emphasized in refs.~\cite{yamazaki2,yamazaki3} that the essential ingredient in KNC's is the 
$\Lambda^* = K^-p$. The strong binding force in KNC's originates not only from the direct $\bar{K}N$ interaction, 
but also from the exchange integral arising from the "Platz-Wechsel" (place-exchange) effect a la Heitler-London 
type mechanism~\cite{HL} for hydrogen molecular bonding. This multi-body attraction was named ``super-strong 
nuclear force"~\cite{yamazaki2}.
 
Parallel to the theoretical activities, experimental searches for KNC's have been carried 
out, but so far, most of the trials are not conclusive. The FINUDA group at DAPHNE first 
reported a $K^-pp$-like peak in the invariant-mass spectrum of $\Lambda-p$ that were 
emitted in $K^-$ capture by light targets~\cite{FINUDA}, but this result was poor in 
statistics, and moreover, its interpretation of the observed spectrum in terms of a 
single Lorentzian peak without background component to yield a binding energy of $B_K = 115 \pm 7$ MeV 
and a width of $\Gamma=67\pm14$ MeV was questioned~\cite{Ramos}. 

In 2007 a theoretical study of the structure of $K^-pp$ and its formation in the 
$d (\pi^+, K^+)$ reaction and in the $p+p\rightarrow{K}^{+}+K^{-}pp$ reaction was 
performed~\cite{yamazaki3}. The former method followed a well-known hypernuclear 
formation, but the formation probability of $K^-pp$ was calculated to be about 1 \% 
as much as the quasi-free background component. With such a pessimistic prediction 
and the non-availability of a suitable beam line and detection system no experimental 
trial had been challenged untill a recent J-PARC E27 experiment~\cite{e27}. Concerning 
the other method using the $p + p$ reaction, a very exotic formation mechanism was 
theoretically revealed in contrast to the conventional pessimistic expectation. In 
such a high-energy collision a large momentum around 1.6 GeV/c is transferred to the 
formed system, and thus, the sticking of $K^-$ to the involved nucleus should be 
enormously small. On the contrary to the pessimistic view, the calculated cross 
section for $K^{-}pp$ was found to be as  large as the free production of $\Lambda^{*}=\Lambda$(1405). 
The reason for this surprising  paradoxical consequence is that the formed state $K^-pp$ 
is a condensed object in which $\Lambda^{*}$ and $p$ are bound with high internal momenta, 
which can be populated by high-energy short-range collisions of $p+p$. The produced $\Lambda^*$ 
is in the short proximity of the participating proton in the collision. A small working group 
(M. Maggiora, K. Suzuki, P. Kienle and T.Yamazaki) was formed to examine this surprising hypothesis 
using large amounts of existing exclusive data of $p+p\rightarrow p+\Lambda+K^{+}$ reactions, 
taken by the DISTO collaboration at Saturne of Saclay. In the conventional view, where the 
$K^-pp$ is not dense, no such reaction will take place. Only when the $K^{-}pp$ were unusually 
dense, a peak comparable to the free emission of $\Lambda^*$ would be observed. In 2010 
the DISTO group published the discovery of a gigantic peak in~\cite{yamazaki4} using the data 
at the incident energy of $T_p=$ 2.85 GeV. Its mass was found to be $M_X=2267\pm 2 (stat)\pm 5 (syst)$ 
MeV/$c^2$, and a binding energy of $B_X=$ 105 MeV and a width of $\Gamma_X=118\pm 8 (stat)\pm 10 (cyst)$ 
MeV were deduced. Recently, another report on the same reaction, but with an incident energy of 
2.5 GeV was reported by the same group~\cite{Kienle}. The observed absence of the peak X at the 
$T_p=$ 2.5 GeV was interpreted as being due to the incident proton energy too low to produce 
the $\Lambda^*$ doorway. More recently, the HADES group at GSI reported absence of X at the 
incident energy of 3.5 GeV~\cite{HADES}. This was interpreted to be due to the too high incident 
energy, which made the collision dynamics to sit outside the favorite Dalitz zone of double 
resonance that was realized at $T_p$.    

We believe it to be vitally important to extend the theoretical and experimental search 
to four-body KNC's. In the present study, we solve the Alt-Grassberger-Sandhas (AGS) equations 
for $\bar{K}NN$ and $\bar{K}NNN$ with an early phenomenological model of $\bar{K}N$ 
interaction by applying our approach based on the coupled-channel AGS equations developed 
in~\cite{shev2,fix}.

This paper is composed as follows. In sect. \ref{formal}, we first give a brief recapitulation 
of the three-body equations and then present the formula corresponding to the four-body equations. 
The inputs for the AGS system of equations are given in sect. \ref{inp}. A discussion of the 
results can be found in section \ref{result}. Finally, we summarize our conclusions in sect. \ref{conclu}.
%%%%%%%%%%%%%%%%%%%%%%%%%%%%%%%%%%%%%%%%%%%%%%%%%%%%%%%%%%%%%%%%%%%%%%%%%%%%%%%%%%%%%%%%%%%%%%%%%%%
\section{Formulation of the problem}
\label{formal}
%%%%%%%%%%%%%%%%%%%%%%%%%%%%%%%%%%%%%%%%%%%%%%%%%%%%%%%%%%%%%%%%%%%%%%%%%%%%%%%%%%%%%%%%%%%%%%%%%%%
\subsection{Three-body AGS equations}
%%%%%%%%%%%%%%%%%%%%%%%%%%%%%%%%%%%%%%%%%%%%%%%%%%%%%%%%%%%
In the present work, we employ the three- and four-body Faddeev equations in momentum space, using the 
Alt-Grassberger-Sandhas form~\cite{alt}. Three-body Faddeev equations~\cite{shev2} in the AGS 
form are given by
\begin{equation}
\mathcal{K}_{ij,I_{i} I_{j}}^{\alpha\beta}=\delta_{\alpha\beta}
\mathcal{M}_{ij,I_{i}I_{j}}^{\alpha\beta}
+\sum_{k,I_{k};\gamma}\mathcal{M}_{ik,I_i I_k}^{\alpha}
\tau_{k,I_k}^{\alpha\gamma}
\mathcal{K}_{kj,I_k I_j}^{\gamma\beta},
\label{ags1}
\end{equation}
where the operator $\mathcal{K}_{ij,I_{i} I_{j}}^{\alpha\beta}$ is the transition amplitude between 
channels $\alpha$ and $\beta$, the operator $\mathcal{M}_{ij,I_{i}I_{j}}^{\alpha\beta}$ is the corresponding 
Born term and $\tau_{i,I_i}^{\alpha\beta}$ is the two-body t-matrix embedded in three-body system. 
Here, the Faddeev partition indices $i,j=$ 1, 2, 3 denote simultaneously a spectator particle and, an 
interacting pair while the particle indices $\alpha,\beta=$ 1, 2, 3 denote the three-body channels. We 
use these Faddeev equations to solve the $\bar{K}NN-\pi\Sigma{N}$ three-body system. Depending on the 
two nucleon spin and isospin, we should treat the $K^{-}pp$ or $K^{-}d$ systems. 
The calculation scheme, which formally allows an exact solution, is based on the separable approximation 
of the appropriate integral kernels. The separable approximation of the kernel of the Faddeev integral 
equation permits one to represent the dynamical equations in terms of particle exchange diagrams~\cite{fix}. 
The key ingredient of the quasi-particle method~\cite{alt2,nadro} is the separable representation of 
the off-shell scattering amplitudes for the two- and three-body systems. We have to introduce also 
the separable representation for the three-body amplitudes and driving terms, which will be necessary 
to find the pole position of $\bar{K}NN$ system. For this purpose we apply the Hilbert-Schmidt expansion (HSE) method
\begin{equation}
\mathcal{M}_{ij,I_i I_j}^{\alpha}(p,p',\epsilon)=-\sum^{N_{r}}_{n=1}\lambda_{n}(\epsilon)
u_{n;i,I_i}^{\alpha}(p,\epsilon)u_{n;j,I_j}^{\alpha}(p',\epsilon),
\label{ags}
\end{equation}
where the form factors $u_{n;i,I_i}^{\alpha}(p,\epsilon)$ are taken as the eigenfunctions of the kernel 
of eq. (\ref{ags1}), with the eigenvalues $\lambda_{n}(\epsilon)$.

The separable form of the Faddeev transition amplitudes is given by
\begin{equation}
\mathcal{K}_{ij,I_i I_j}^{\alpha\beta}(p,p',\epsilon)=\sum^{N}_{n=1}u_{n;i,I_i}^{\alpha}
(p,\epsilon)\zeta_{n}(\epsilon)u_{n;j,I_j}^{\beta}(p',\epsilon),
\label{ags2}
\end{equation}
where the functions $\zeta_{n}(\epsilon)$ obey the equation
\begin{equation}
\zeta_n(\epsilon)=\lambda_n(\epsilon)/(\lambda_n(\epsilon)-1).
\label{zet}
\end{equation}

Then using the separable approximation for the Faddeev amplitudes and driving terms in (\ref{ags1}), 
the Faddeev equations take the form
\begin{equation}
u_{n;i,I_i}^{\alpha}=\frac{1}{\lambda_n}\sum_{k=1}^{3}\sum_{\gamma=1}^{3}\sum_{I_k}
\mathcal{M}_{ik,I_i I_k}^{\alpha}\tau_{k,I_k}^{\alpha\gamma}u_{n;k,I_k}^{\gamma}.
\label{ags3}
\end{equation} 

The AGS equation of (\ref{ags3}) is a Fredholm type integral equation. To find the resonance energy of the 
three-body system using these equations, we should transform the integral equations into algebraic ones and 
then search for a complex energy at which the first eigenvalue of the kernel matrix becomes equal to one. 
Before we proceed to solve the AGS equations for both $(\bar{K}NN)_{s=0,1}$ systems, the operators involving 
two identical baryons should be antisymmetric. The baryon spins do not enter explicitly in the three-body 
equations because the total spin $s$ remains unchanged in the process. In the $K^{-}d$ case, the spin component 
is symmetric, then all operators in isospin base should be antisymmetric. In the case of $K^{-}pp$ the spin 
component is antisymmetric. Thus, all operators in isospin base should be symmetric. 
%%%%%%%%%%%%%%%%%%%%%%%%%%%%%%%%%%%%%%%%%%%%%%%%%%%%%%%%%%%%%%%%%%%%%%%%%%%%%%%%%%%%%%%%%%%%%%%%%%%
\subsection{The four-body $\bar{K}NNN$ equations}
\begin{figure}[htb]
%\vspace*{4cm} 
\begin{center}
\resizebox{0.6\textwidth}{!}{%
\includegraphics{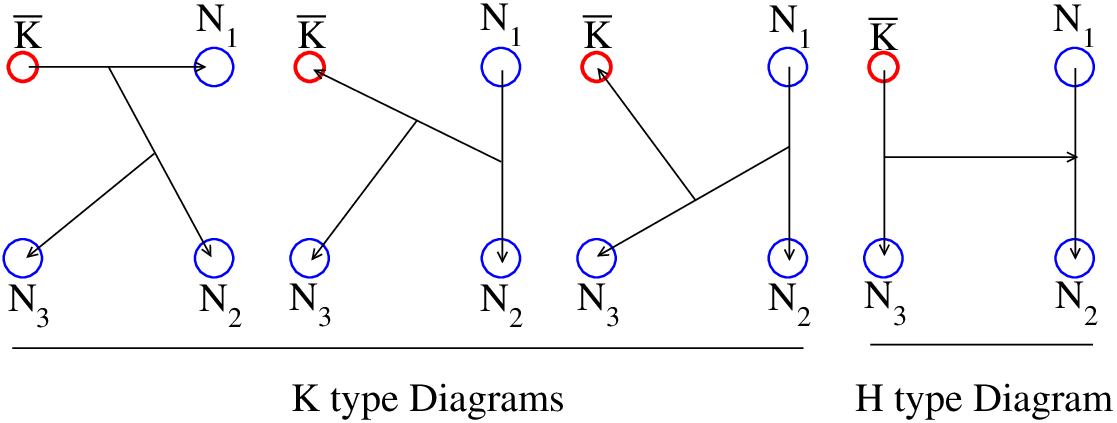}}
%\vspace*{-3.5cm}
\end{center}
\caption{The four different rearrangement channels of the $\bar{K}NNN$ four-body system including 
the K- and H-type diagrams are represented. Antisymmetrization of three $N$'s is to be made within each channel.}
\label{kh}
\end{figure}

%%%%%%%%%%%%%%%%%%%%%%%%%%%%%%%%%%%%%%%%%%%%%%%%%%%%%%%%%%%%%%%%%%%%%%%%%%%%%%%%%%%%%%%%%%%%%%%%%%%
In four-body $\bar{K}NNN$ system, there is three identical nucleons, therefore, the four-body equations for 
$\bar{K}NNN$ system are reduced to three sets of integral equations. As it is shown in fig. \ref{kh}, the whole 
dynamics is described in terms of the Faddeev amplitudes, which connect the three channels characterized by the 
following partitions
\begin{equation}
\alpha=\{1,2,3\}=\{\bar{K}(NNN),N(\bar{K}NN),(\bar{K}N)(NN)\}.
\label{chan}
\end{equation}
%%%%%%%%%%%%%%%%%%%%%%%%%%%%%%%%%%%%%%%%%%%%%%%%%%%%%%%%%%%
%%%%%%%%%%%%%%%%%%%%%%%%%%%%%%%%%%%%%%%%%%%%%%%%%%%%%%%%%%%

We need all possible amplitudes connecting the initial state, consisting of the 3$N$ bound state 
($\mathrm{^{3}{He}}$) and a free kaon, with all three channels listed in (\ref{chan}) via particle or 
two-body quasi-particle exchange. The four-body Faddeev amplitudes obey a set of three coupled integral 
equations, whose structure is represented by the following matrix equation

\begin{equation}
\begin{pmatrix} {\mathcal{A}}^{11}  \\ {\mathcal{A}}^{21} \\ {\mathcal{A}}^{31} \end{pmatrix}=
\begin{pmatrix} 0 & \mathcal{R}^{12} & \mathcal{R}^{13} \\ \mathcal{R}^{21} & \mathcal{R}^{22} 
& \mathcal{R}^{23} \\ \mathcal{R}^{31} & \mathcal{R}^{32} & 0 \end{pmatrix}
\begin{pmatrix} \zeta^1 & 0 & 0 \\ 0 & \zeta^2 & 0 \\ 0 & 0 & \zeta^3 \end{pmatrix}
\begin{pmatrix} {\mathcal{A}}^{11}  \\ {\mathcal{A}}^{21} \\ {\mathcal{A}}^{31} \end{pmatrix}.
\label{rotation_matrix} 
\end{equation}

Here, we take into account only the dominant s-wave part of the interaction in the two-body subsystems and 
thus in the three- and four-particle states. Therefore, in all expressions, we drop the index $L=0$. The 
explicit analytical form of the transition amplitudes between the channel states, taking into account the 
spin and isospin degrees of freedom, are given by
\begin{equation}
\mathcal{A}^{\alpha\beta,ss'}_{II',nn'}=\mathcal{R}^{\alpha\beta,ss'}_{II',nn'}+
\sum^{3}_{\gamma=1}\sum_{n''s''I''}\mathcal{R}^{\alpha\gamma,ss''}_{II'',nn''}
\zeta^{\gamma}_{n''}\mathcal{A}^{\gamma\beta,s''s'}_{I''I',n''n'},
\label{trans1}
\end{equation}
where the operators $\mathcal{A}^{\alpha\beta,ss'}_{II',nn'}$ are the four-body Faddeev amplitudes, 
$\zeta^{\gamma}_{n}$-functions are represented by eq. (\ref{zet}) and the operators 
$\mathcal{R}^{\alpha\beta,nn'}_{sI,sI'}$ are driving terms, which describe the effective particle-exchange 
potential realized by the exchanged particle between the quasi-particles in the channels $\alpha$ and $\beta$, 
which can be written as

\begin{eqnarray}
\mathcal{R}^{\alpha\beta,ss'}_{II',nn'}(p,p',E)&=&\frac{\Omega^{ss'}_{II'}}{2}
\int^{+1}_{-1}d(\hat{p}\cdotp\hat{p}'){u}^{\alpha,s}_{n,I}(\vec{q},\epsilon_{\alpha}-
\frac{p^{2}}{2\mathcal{M}_{\alpha}}) \nonumber \\ &\times&
\tau(z)u^{\beta,s'}_{n',I'}(\vec{q'},\epsilon_{\beta}-\frac{p'^{2}}{2\mathcal{M}_{\beta}}).
\label{trans2}
\end{eqnarray}

Here, the symbols $\Omega^{ss'}_{II'}$ are the spin and isospin Clebsch-Gordan coefficients, 
the functions ${u}^{\alpha,s}_{n,I}$ are the form factors that generated by the separable 
representation of the sub-amplitudes appearing in the channels (\ref{chan}) and $z$ is given as 
$z=E-\frac{p^{2}}{2M_{\beta}}-\frac{p'^{2}}{2M_{\alpha}}-\frac{\vec{p}\cdot\vec{p}'}{m}$.
The energy $\epsilon_{\alpha}$ is the subsystem energy in channel $\alpha$. The momenta 
$\vec{q}(\vec{p},\vec{p}')$ and $\vec{q}'(\vec{p},\vec{p}')$ are given in terms of $\vec{p}$ 
and $\vec{p'}$. We use the relations

\begin{equation}
\begin{split}
& \vec{q}=\vec{p}'+\frac{M_{\alpha}}{m}\vec{p}, \\
& \vec{q}'=\vec{p}+\frac{M_{\beta}}{m}\vec{p}',
\end{split}
\label{trans3}
\end{equation}
where $m$ is exchanged particle or quasi-particle mass and the reduced masses $\mathcal{M}_{\alpha}$
and $M_{\alpha}$ in the channel $\alpha$ of the [3+1] subsystem are defined by
\begin{equation}
\begin{split}
& \mathcal{M}_{\alpha} = m^{\alpha}_{i}(m^{\alpha}_{j}+m^{\alpha}_{k}+m^{\alpha}_{l})
/(m^{\alpha}_{i}+m^{\alpha}_{j}+m^{\alpha}_{k}+m^{\alpha}_{l}), \\
& M_{\alpha} = m^{\alpha}_{j}(m^{\alpha}_{k}+m^{\alpha}_{l})/(m^{\alpha}_{j}+m^{\alpha}_{k}+m^{\alpha}_{l}),
\end{split}
\label{trans4}
\end{equation}
and in the case of the [2+2] subsystem are given by
\begin{equation}
\begin{split}
& \mathcal{M}_{\alpha}=(m^{\alpha}_{i}+m^{\alpha}_{j})(m^{\alpha}_{k}+m^{\alpha}_{l})
/(m^{\alpha}_{i}+m^{\alpha}_{j}+m^{\alpha}_{k}+m^{\alpha}_{l}), \\
& M_{\alpha} = m^{\alpha}_{i}m^{\alpha}_{j}/(m^{\alpha}_{i}+m^{\alpha}_{j}).
\end{split}
\label{trans44}
\end{equation}

\begin{figure*}
%\vspace*{6cm} 
\begin{center}
\resizebox{0.8\textwidth}{!}{%
\includegraphics{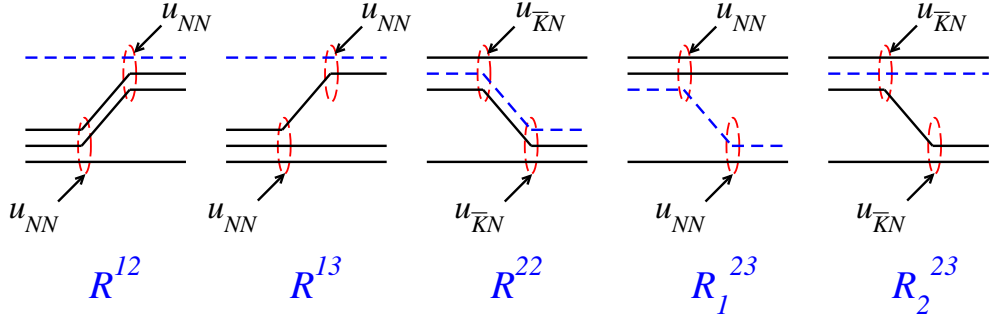}}
%\vspace*{-6.cm}
\end{center}
\caption{Diagrammatic representation of the potentials $\mathcal{R}^{\alpha\beta}$ in the separable 
approximation. The blue dashed line corresponds to the $\bar{K}$ and the black solid lines correspond 
to the nucleon. The symbols $u_{\alpha}$ will define the initial and final state of the system.}
\label{diag}
\end{figure*}

The meaning of the driving terms $\mathcal{R}^{\alpha\beta,ss'}_{II',nn'}$ is explained schematically by the 
diagrammatic representation in fig. \ref{diag}. By cyclic permutation of the nucleons, one can obtain various 
relations between the different driving terms $\mathcal{R}^{\alpha\beta,ss'}_{II',nn'}$. For example, by applying a 
combination of a cyclic permutation within an antisymmetrized $NN$-state, one obtains for the transition 
$2\rightarrow{3}$ the relation
\be
\mathcal{R}^{23}=\mathcal{R}_{1}^{23}+2\mathcal{R}_{2}^{23},
\ee
where the coefficient 2 in the term $\mathcal{R}_{2}^{23}$ comes from the identity of the nucleons. 
%%%%%%%%%%%%%%%%%%%%%%%%%%%%%%%%%%%%%%%%%%%%%%%%%%%%%%%%%%%
%%%%%%%%%%%%%%%%%%%%%%%%%%%%%%%%%%%%%%%%%%%%%%%%%%%%%%%%%%%

Before we proceed to solve the four-body equations, we also need as input the equations describing two independent 
pairs of interacting particles $(\bar{K}N)+(NN)$. The corresponding equations read in our case

\begin{equation}
\begin{split}
& \mathcal{Y}^{sI,s'I'}_{\bar{K}N,NN}=\mathcal{W}^{sI,s'I'}_{\bar{K}N,NN}+\mathcal{W}^{sI,s'I'}_{\bar{K}N,NN}
\tau^{s'I'}_{NN}\mathcal{Y}^{s'I',s'I'}_{NN,NN}, \\
& \mathcal{Y}^{s'I',s'I'}_{NN,NN}=\mathcal{W}^{s'I',sI}_{NN,\bar{K}N}\tau^{sI}_{\bar{K}N}\mathcal{Y}^{sI,s'I'}_{\bar{K}N,NN}.
\end{split}
\label{trans5}
\end{equation}

\begin{figure}[htb]
%\vspace{2.cm}
\begin{center}
\resizebox{0.7\textwidth}{!}{%
\includegraphics{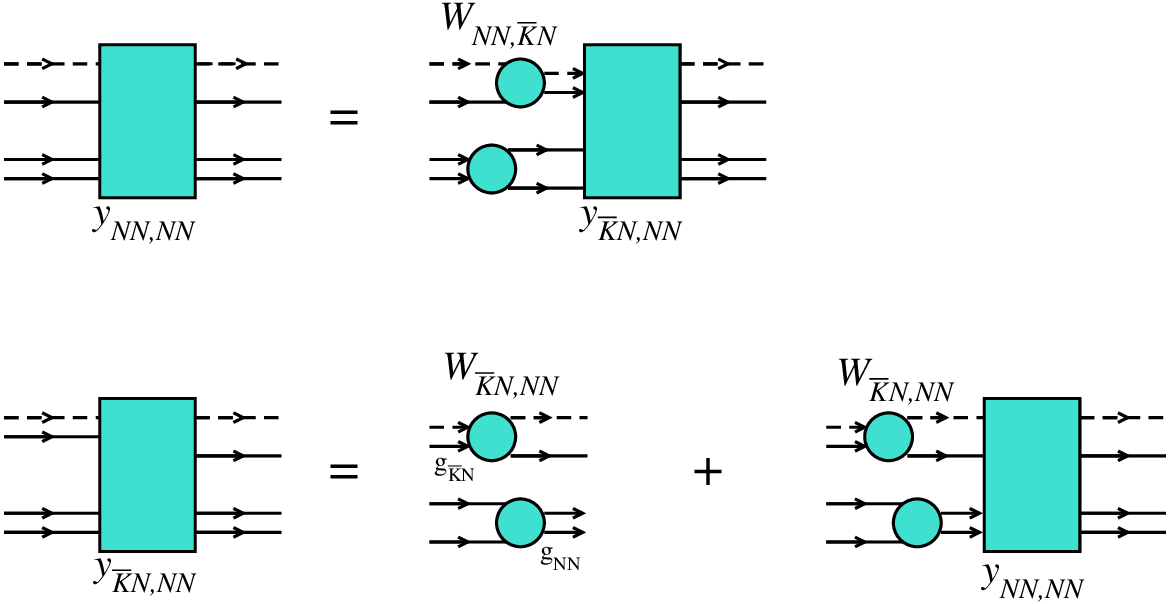}}
\end{center}
%\vspace{-2cm}
\caption{Diagrammatic representation of the equation (\ref{trans5}) for the transition amplitudes 
$\mathcal{Y}^{sI,s'I'}_{i,j}$ of $(\bar{K}N)-(NN)$ system. The symbols $g_{\bar{K}N}$ and $g_{NN}$ 
are the form factors of the $\bar{K}N$ and $NN$ interactions.}
\label{htype}
\end{figure}

Here, the operators $\mathcal{Y}^{sI,s'I'}_{i,j}$ are the Faddeev amplitudes which describe two independent 
pairs of interacting particles and the operators $\mathcal{W}^{sI,s'I'}_{i,j}$ are the effective potentials. 
A graphical representation of the system (\ref{trans5}) is shown in fig. \ref{htype}. Analogously to the 
treatment in the previous subsection, the separable form of the amplitude can easily be found
\begin{equation}
\mathcal{Y}_{i,j}^{sI,s'I'}(p,p',\epsilon)=\sum^{N_{r}}_{n=1}u_{n;i}^{sI}(p,\epsilon)\zeta_{n}(\epsilon)u_{n;j}^{s'I'}(p',\epsilon),
\label{trans6}
\end{equation}
where the functions $u_{n;i}^{sI}$ are the eigenfunctions of the kernel of eq. (\ref{trans5}).
\begin{equation}
u_{n;i}^{sI}=\frac{1}{\lambda_n}\sum_{j=\bar{K}N,NN}
\mathcal{W}^{sI,s'I'}_{i,j}\tau^{s'I'}_{j}u_{n;j}^{s'I'}.
\label{trans7}
\end{equation}

The conversion of the four-body equations to a numerically manageable form is yielded by expanding the 
two- and three-body Faddeev amplitudes in eqs. (\ref{ags1}) and (\ref{trans5}) into separable series of finite 
rank $N_{r}$. For to make a separable representation for these subsystem amplitudes, one can use the energy dependent 
pole expansion (EDPE)~\cite{sofia} or the Hilbert-Schmidt expansion~\cite{nadro}. The desired approach in this 
work is the Hilbert-Schmidt expansion (HSE). The inputs for the driving terms of equation (\ref{trans2}) are two-body 
t-matrices, embedded in the four-body Hilbert space and the form factors, which are defined in eqs. (\ref{ags3}) 
and (\ref{trans7}). Before we proceed to solve the AGS equations (\ref{trans1}), we should antisymmetriz the 
basic amplitudes with respect to the exchange of the nucleons for which we follow mainly the work of~\cite{fix}.
%%%%%%%%%%%%%%%%%%%%%%%%%%%%%%%%%%%%%%%%%%%%%%%%%%%%%%%%%%%
%%%%%%%%%%%%%%%%%%%%%%%%%%%%%%%%%%%%%%%%%%%%%%%%%%%%%%%%%%%
%%%%%%%%%%%%%%%%%%%%%%%%%%%%%%%%%%%%%%%%%%%%%%%%%%%%%%%%%%%%%%%%%%%%%%%%%%%%%%%%%%%%%%%%%%%%%%%%%%%
\section{Two-body interactions}
\label{inp}
%%%%%%%%%%%%%%%%%%%%%%%%%%%%%%%%%%%%%%%%%%%%%%%%%%%%%%%%%%%%%%%%%%%%%%%%%%%%%%%%%%%%%%%%%%%%%%%%%%%
All two-body interactions are taken in $s$-wave and separable form. Thus, in the case of separable 
two-body potential we have
\be
V_{\alpha\beta}(p_{\alpha},p_{\beta})=\lambda_{\alpha\beta}g_{\alpha}(p_{\alpha})g_{\beta}(p_{\beta}).
\ee

Here, $\alpha$ and $\beta$ enumerate two-body channels and $p_{\alpha}$ is the c.m. momentum in the 
corresponding channel. The two-body t-matrices that serve as input for the three- and four-body problem are all taken 
in the separable form for a given partial wave
\be
T_{\alpha\beta}(p_{\alpha},p_{\beta},E)=g_{\alpha}(p_{\alpha})\tau_{\alpha\beta}(E)g_{\beta}(p_{\beta}),
\ee
where $E$ is the total energy, $\lambda_{\alpha\beta}$ are the coupling strength parameters of the 
interaction and the form factors are defined by $g_{\alpha}(p_{\alpha})$. 

The $\bar{K}N$ interaction, which is the most important interaction for the $\bar{K}NN$ and $\bar{K}NNN$ 
systems, is usually described either by pure phenomenological or by chirally motivated potentials. In 
our Faddeev calculations, we use two different effective interactions for the coupled-channel $\bar{K}N-\pi\Sigma$ 
interaction that, having a one- and two-pole structure of the $\Lambda$(1405) resonance. The potentials that 
we use here for the $\bar{K}N$ interaction are given in ref.~\cite{shev4}. 
The parameters of the coupled-channel $\bar{K}N-\pi\Sigma$ potential were fitted to reproduce all existing 
experimental data on the low-energy $\bar{K}N$ system and the fitting was performed by using physical masses 
in $\bar{K}N$ and $\pi\Sigma$ channels with the inclusion of the Coulomb interaction.

The $s$-wave $\Sigma{N}$ interaction in the $I=1/2$ isospin state is coupled with $\Lambda{N}$ channel, 
therefore, we used an optical potential for $\Sigma{N}$ interaction in this isospin state and a real 
potential for $I=3/2$ channel. The parameters chosen for the $\Sigma{N}$ interaction were those given in 
ref.~\cite{shev5}. In this calculation, we use the spin independent version of $\Sigma{N}$ interaction.

In our three- and four-body study for singlet and triplet $NN$ interaction, we choose a potential of PEST 
type~\cite{para}, which is a separablization of the Paris potential. The coupling strength parameter was set to 
$\lambda=-1$ and the form factors are defined by
\be
g^{NN}_{s,I}(p)=\frac{1}{2\sqrt{\pi}}\sum^{6}_{n=1}\frac{c^{NN}_{n,I}}{p^2+(\beta^{NN}_{n,I})^2},
\ee
where the constants $c^{NN}_{n,I}$ and $\beta^{NN}_{n,I}$ are listed in ref.~\cite{para}. PEST potential is 
equivalent to the Paris potential for energies up to $E_{lab}\sim50$ MeV. It reproduces the deuteron binding energy 
$E_{B.E}=2.2249$ MeV, as well as the singlet and triplet $NN$ scattering lengths, $a(^{1}{S}_{0})=17.534$ fm and 
$a(^{3}{S}_{1})=−5.422$ fm, respectively. The $\mathrm{^{3}{He}}$ binding energy, calculated with PEST potential is 
$9.7$ MeV while the experimental value is $8.54$ MeV.
%%%%%%%%%%%%%%%%%%%%%%%%%%%%%%%%%%%%%%%%%%%%%%%%%%%%%%%%%%%%%%%%%%%%%%%%%%%%%%%%%%%%%%%%%%%%%%%%%%%
\section{Results and discussions}
\label{result}
\begin{figure}[H]
\begin{center}
\centering
\includegraphics[scale=0.4]{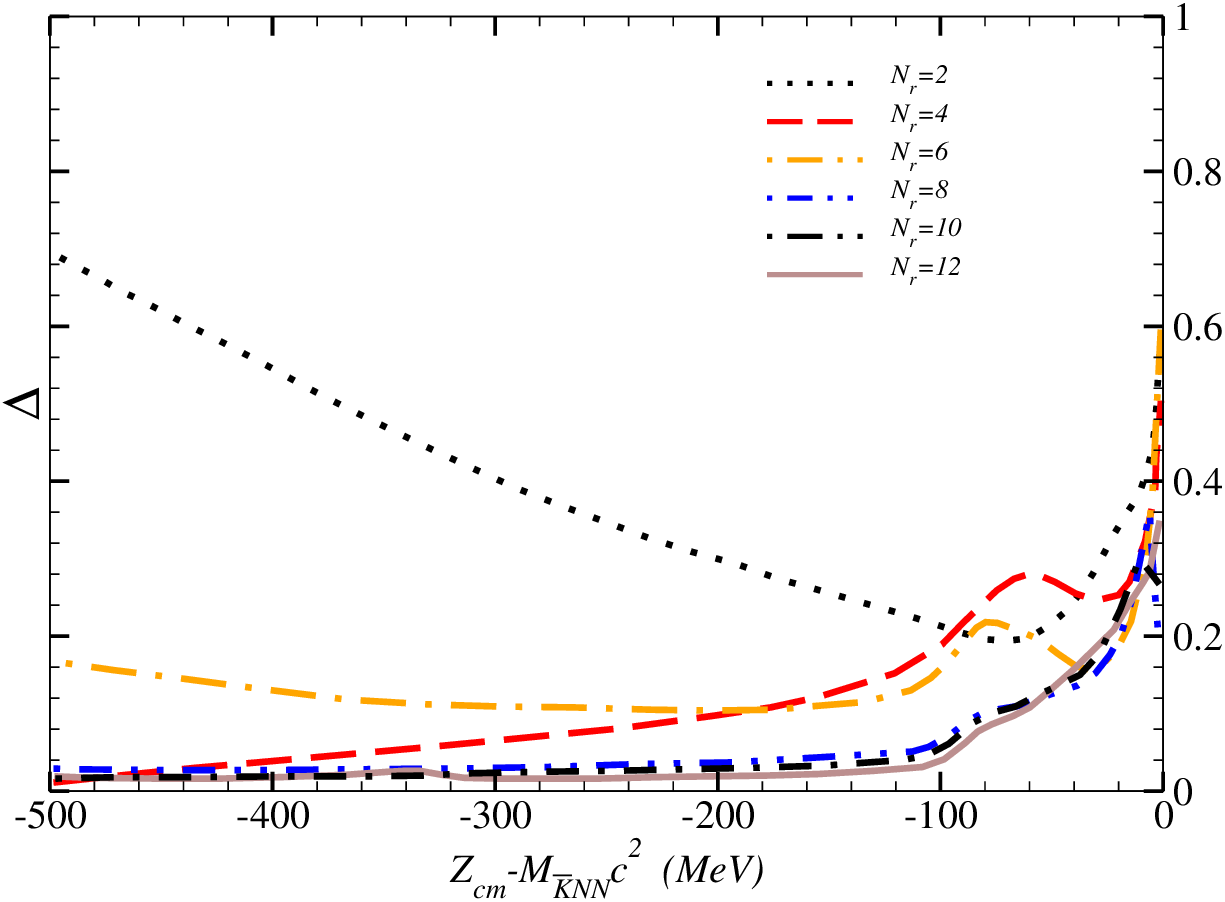}
\end{center}
\caption{(Color online) The ratio between the Schmidt norms of the kernels $\vartheta$ and $\vartheta_{N_{r}}$ as defined 
by eqs. (\ref{ratio}) and (\ref{pot}).}
\label{conver}
\end{figure}

%%%%%%%%%%%%%%%%%%%%%%%%%%%%%%%%%%%%%%%%%%%%%%%%%%%%%%%%%%%%%%%%%%%%%%%%%%%%%%%%%%%%%%%%%%%%%%%%%%%
Because $[\bar{K}NN]_{I=1/2,J^{\pi}=0^{-}}$ is the most important subsystem of the four-body $\bar{K}NNN$ 
system, in fig.~\ref{conver} we demonstrated how well a finite sum (\ref{ags}) may represent the exact 
amplitude. Thus, we calculated the ratio of the Schmidt norm for 
\begin{equation}
\varDelta=\frac{\|\vartheta_{N_{r}}\|}{\|\vartheta\|},
\label{ratio}
\end{equation}
of the operators
\begin{equation}
\begin{split}
& \vartheta=\mathcal{M}_{(\bar{K}N)_{I=0}N-(\bar{K}N)_{I=0}N},\\
& \vartheta_{N_{r}}=\mathcal{M}_{(\bar{K}N)_{I=0}N-(\bar{K}N)_{I=0}N}-\mathcal{M}^{N_{r}}_{(\bar{K}N)_{I=0}N-(\bar{K}N)_{I=0}N},
\end{split}
\label{pot}
\end{equation}
where $\mathcal{M}^{N_{r}}_{(\bar{K}N)_{I=0}N-(\bar{K}N)_{I=0}N}$ is given by the sum (\ref{ags}) containing only the 
first $N_{r}$ terms. One can see that the rate of convergence is not very effective, but appears to be sufficient for 
the practical calculation.
%%%%%%%%%%%%%%%%%%%%%%%%%%%%%%%%%%%%%%%%%%%%%%%%%%%%%%%%%%%
%%%%%%%%%%%%%%%%%%%%%%%%%%%%%%%%%%%%%%%%%%%%%%%%%%%%%%%%%%%
As a starting three- and four-body calculation, we calculated the binding energies and widths of $K^{-}pp$ 
and $K^{-}ppn$ quasi-bound states using a one-channel complex $\bar{K}N$ potential~\cite{shev2}. 
During these calculations, we considered the $\bar{K}N$ potentials with the parameters $\lambda^{I,Complex}_{\bar{K}N,
\bar{K}N}$ and $\beta_{I}$, which reproduce $\mathrm{M}_{\Lambda}=$ 1405.1 MeV, $\Gamma_{\Lambda}=$ 50 MeV and 
the $K^{-}p$ scattering length, for which we used as a guideline the SIDDHARTA measured value: 
$a^{\mathrm{SIDD}}_{K^{-}p}=(-0.65+i0.81)$ fm~\cite{bazi}. In table \ref{sn1}, our results for the binding 
energy of the $K^{-}pp$ and $K^{-}ppn$ related to these data, and using $\beta_{I}=3.5$ $\mathrm{fm}^{-1}$, 
are represented. In table \ref{sn1} we performed a calculation for the one-channel $\bar{K}NN$ system using 
a one-channel complex $\bar{K}N$ potential. For these data, we found a quasi-bound state for $K^{-}pp$ and 
$K^{-}ppn$ below the threshold.
%%%%%%%%%%%%%%%%%%%%%%%%%%%%%%%%%%%%%%%%%%%%%%%%%%%%%%%%%%%
\begin{table}[H]
\caption{The binding energies and widths of the quasi-bound state of the 
$K^{-}pp$ and $K^{-}ppn$ systems for one-channel complex potential.}
\centering
\begin{tabular}{ccc}
\hline\noalign{\smallskip}
$a_{K^{-}p}$ (fm) & $E_{K^{-}pp}$ (MeV) & $E_{K^{-}ppn}$ (MeV) \\
\noalign{\smallskip}\hline\noalign{\smallskip}
\, -0.65+i0.81~\cite{bazi} \, & \, -49.4-i43.5 \, & \, -60.2-i42.2 \,  \\
\noalign{\smallskip}\hline
\end{tabular}
\label{sn1} 
\end{table}
%%%%%%%%%%%%%%%%%%%%%%%%%%%%%%%%%%%%%%%%%%%%%%%%%%%%%%%%%%%

In the following we present the results for the binding energy of the $\bar{K}N$ and $\bar{K}NN$ in table \ref{sn2} 
for the one- and two-pole version of $\bar{K}N-\pi\Sigma$ interaction. The binding energies for $\bar{K}N$ in 
table \ref{sn2} are just a bit different from those given in the original ref.~\cite{shev4}. The reason is that the 
above calculations were performed with averaged masses and without Coulomb interaction while the fitting to the 
experimental data was performed with physical masses and Coulomb interaction. At the beginning, we solved eq. (\ref{ags3}) 
with neglecting the $\Sigma{N}$ and $\pi{N}$ interactions. Thus, only $\bar{K}N$ and $NN$ t-matrices enter the equations. 
Therefore, we constructed the exact optical $\bar{K}N(-\pi\Sigma)$ potential, which is an approximation for the full 
coupled-channel interaction. The binding energies are calculated with respect to the $\bar{K}NN$ threshold. In the 
third column of table \ref{sn2} the binding energy and width of the full coupled-channel calculation of the 
$\bar{K}NN-\pi\Sigma{N}$ by taking the $\Sigma{N}$ interaction into account are presented. One can see that the 
one-channel AGS calculation with exact optical $\bar{K}N(-\pi\Sigma)$ potential gives a good approximation to the 
full coupled-channel calculations. This result was expected because the exact optical potential provides exactly 
the same elastic $\bar{K}N-\bar{K}N$ amplitude as the coupled-channel model of interaction, see ref.~\cite{shev5}.
%%%%%%%%%%%%%%%%%%%%%%%%%%%%%%%%%%%%%%%%%%%%%%%%%%%%%%%%%%%
\begin{table}[H]
\caption{The sensitivity of the binding energies and widths of the quasi-bound state of the $K^{-}pp$ systems 
to the $\bar{K}N$, $\Sigma{N}$ interactions. $E^0$ stands for no $\Sigma{N}$ interaction, while in calculating 
the $E^1$, $\Sigma{N}$ interaction is on. The real part of the pole $E_{K^{-}pp}$ is measured from the $\bar{K}NN$ threshold.}
\centering
\begin{tabular}{cccc}
\hline\noalign{\smallskip}
& $E_{\bar{K}N}$ (MeV) & $E^{(0)}_{\bar{K}NN}$ (MeV) & $E^{(1)}_{\bar{K}NN}$ (MeV) \\
\noalign{\smallskip}\hline\noalign{\smallskip}
 $V^{SIDD}_{One-pole}$  &  1428.1-i46.6  &  -48.7-i34.3  &  -52.8-i31.5  \\
 $V^{SIDD}_{Two-pole}$  &  1418.1-i56.9  &  -45.4-i24.4  &  -47.1-i25.0  \\
                        &  1382.0-i104.2 &               &               \\
\noalign{\smallskip}\hline
\end{tabular}
\label{sn2} 
\end{table}
%%%%%%%%%%%%%%%%%%%%%%%%%%%%%%%%%%%%%%%%%%%%%%%%%%%%%%%%%%%
%%%%%%%%%%%%%%%%%%%%%%%%%%%%%%%%%%%%%%%%%%%%%%%%%%%%%%%%%%%
\begin{table*}[t]
\caption{The sensitivity of the binding energies and widths of the quasi-bound state of the $K^{-}ppn$ 
system to the number of terms $N_{r}$ in eqs. (\ref{ags}) and (\ref{trans6}). $E^{SIDD,One-pole}_{K^{-}ppn}$ and 
$E^{SIDD,Two-pole}_{K^{-}ppn}$ correspond to the one- and two-pole version of the $\bar{K}N$ interaction, 
respectively. The real part of the pole $E_{K^{-}ppn}$ (in MeV) is measured from the $\bar{K}NNN$ threshold.}
\centering
\begin{tabular}{cccc}
\hline\noalign{\smallskip}
& \, $N_{r}=2$ \, & \, $N_{r}=4$ \, & \, $N_{r}=6$ \, \\
\noalign{\smallskip}\hline\noalign{\smallskip}
$E^{SIDD,One-pole}_{K^{-}ppn}$ \, & \, -69.6-i10.5 \, & \, -69.0-i11.1 \, & \, -68.8-i11.0 \, \\
\noalign{\smallskip}
\noalign{\smallskip}
$E^{SIDD,Two-pole}_{K^{-}ppn}$ \, & \, -56.7-i8.6  \, & \, -56.2-i8.8  \, & \, -55.9-i8.8  \, \\
\noalign{\smallskip}\hline
\end{tabular}
\label{sn3} 
\end{table*}
%%%%%%%%%%%%%%%%%%%%%%%%%%%%%%%%%%%%%%%%%%%%%%%%%%%%%%%%%%%
In table \ref{sn3} we presented our results for the $K^{-}ppn$ quasi-bound state obtained by 
keeping a finite number of terms $N_{r}$, in the Hilbert-Schmidt expansion of the amplitudes 
(\ref{ags2}) and (\ref{trans6}). In this table, the rate of convergence of $K^{-}ppn$ binding 
energy is investigated and one can see that the choice $N_{r}=4$ provides rather satisfactory 
accuracy. In the four-body calculation we have neglected any $\Sigma{N}-\Lambda{N}$ and $\pi{N}$ 
interactions. The inclusion of these interactions would increase the number of channels in the 
four-body equations which would lead to much more complex formalism. As mentioned in the previous 
paragraph, the one-channel AGS calculation with exact optical $\bar{K}N$ potential, giving exactly 
the same $\bar{K}N-\bar{K}N$ amplitude as the corresponding coupled-channel potential, turns out 
to be a good approximation. Therefore, one can safely assume that $\Sigma{N}-\Lambda{N}$ and 
$\pi{N}$ interactions in the $\pi\Sigma{NN}$ channel can not change the binding energy of the 
$\bar{K}NNN-\pi\Sigma{NN}$ system more than a few MeV. Using the exact optical $\bar{K}N$ potential, 
our two-channel four-body calculation with coupled-channel $\bar{K}N-\pi\Sigma$ potential will be 
equivalent to the one-channel four-body calculation.

The binding energies and widths of the quasi-bound state of the $K^{-}pp$, $K^{-}ppn$ and $K^{-}ppp$ systems 
have been calculated and presented in table \ref{sn4}. We calculated $K^{-}ppn$ and $K^{-}ppp$ quasi-bound state positions by 
keeping four terms in the Hilbert-Schmidt expansion of the amplitudes (\ref{ags2}) and (\ref{trans6}). 

Very recently, some few-body calculations are performed on $K^{-}ppn$ by the variational method~\cite{gal,roman} and 
the Faddeev approach~\cite{maeda}. The investigation of the $\bar{K}NNN$ in ref.~\cite{gal} uses the effective 
$\bar{K}N$ interaction derived from chiral low energy theorem, a quasi-bound state was found with a binding energy 
30 MeV and a width $30$ MeV below the threshold energy of the $\bar{K}NNN$ state. A similar conclusion was drawn 
using the Faddeev equation by Maeda {\it et al.} using a one-channel real potential~\cite{maeda}. The obtained binding 
energies for $K^{-}ppn$ was about $69$ MeV below threshold energy. The obtained binding energies of the $\bar{K}NNN$ 
quasi-bound state in ref.~\cite{roman} for A-Y and HW potentials are $\sim$ 65 and $\sim$ 18 MeV and the corresponding 
widths are $\sim$ 74-80 and $\sim$ 27-31, respectively. The comparison our results for $K^{-}ppn$ obtained for PEST 
$NN$ interaction and the coupled-channel $\bar{K}N-\pi\Sigma$ interaction with calculations in ref.~\cite{maeda} within 
the Faddeev method for rank-two $NN$ interaction and one-channel real $\bar{K}N$ interaction shows that they are in 
the same range. However, this is in contrast to the chiral low energy potential, which is constructed to generate a 
bound state with a binding energy $\sim$ 30 MeV.
%%%%%%%%%%%%%%%%%%%%%%%%%%%%%%%%%%%%%%%%%%%%%%%%%%%%%%%%%%%
\begin{table}[H]
\caption{Pole positions (in MeV) of the quasi-bound states in the $K^{-}pp$, $K^{-}ppn$ and $K^{-}ppp$. The Faddeev 
AGS calculations performed with the phenomenological potentials from ref.~\cite{shev4}. The potentials $V^{SIDD}_{One-pole}$ 
and $V^{SIDD}_{Two-pole}$ are $\bar{K}N-\pi\Sigma$ potentials, which produce the one- and two-pole structure of the 
$\Lambda$(1405) resonance, respectively. The binding energies (real part of the pole) are measured from the thresholds.}
\centering
\begin{tabular}{cccc}
\hline\noalign{\smallskip}
& $E_{K^{-}pp}$  &  $E_{K^{-}ppn}$  &  $E_{K^{-}ppp}$   \\
\noalign{\smallskip}\hline\noalign{\smallskip}
 $V^{SIDD}_{One-pole}$  &  -48.7-i34.3  &  -68.8-i11.0  &  -99.6-i10.5 \\
\noalign{\smallskip}
 $V^{SIDD}_{Two-pole}$  &  -45.4-i24.4  &  -55.9-i8.8   &  -87.8-i3.5   \\
\noalign{\smallskip}\hline
\end{tabular}
\label{sn4} 
\end{table}
%%%%%%%%%%%%%%%%%%%%%%%%%%%%%%%%%%%%%%%%%%%%%%%%%%%%%%%%%%%
%%%%%%%%%%%%%%%%%%%%%%%%%%%%%%%%%%%%%%%%%%%%%%%%%%%%%%%%%%%%%%%%%%%%%%%%%%%%%%%%%%%%%%%%%%%%%%%%%%%
\section{Conclusion}
\label{conclu}
%%%%%%%%%%%%%%%%%%%%%%%%%%%%%%%%%%%%%%%%%%%%%%%%%%%%%%%%%%%%%%%%%%%%%%%%%%%%%%%%%%%%%%%%%%%%%%%%%%%
Starting from Faddeev AGS equations and using different versions of the $\bar{K}N-\pi\Sigma$ 
potentials, which produce the one- and two-pole structure of the $\Lambda$(1405) resonance and 
separable expressions for the [3+1] and [2+2] subsystems. We employed the HSE method to reduce 
the problem to a set of single-variable integral equations. We solved the three- and four-body 
Faddeev equations, searching for $K^{-}pp$, $K^{-}ppn$ and $K^{-}ppp$ quasi-bound states. We 
studied the dependence of the pole energy on different models of $\bar{K}N-\pi\Sigma$ interaction. 
It was shown that a one-channel complex $\bar{K}N$ potential gives much broader three- and four-body 
quasi-bound state than the exact optical potential. The calculations yielded binding energy 
$B_{K^{-}pp}\sim$ 45-55, $B_{K^{-}ppn}\sim$ 55-70 and $B_{K^{-}ppp}\sim$ 90-100 MeV for $K^{-}pp$, 
$K^{-}ppn$ and $K^{-}ppp$, respectively. The obtained widths for these systems are $\Gamma_{K^{-}pp}\sim$ 50-75, 
$\Gamma_{K^{-}ppn}=16-20$ and $\Gamma_{K^{-}ppp}=7-20$ MeV. However, a similar calculation should 
be performed for the standard energy-dependent $\bar{K}N$ input potential, too. The quasi-bound 
states resulting from the energy-dependent potentials happen to be shallower, this is due to the 
energy dependence of the interaction. The energy-dependent potential will provide a weaker $\bar{K}N$ 
attraction for lower energies than the energy independent potential under consideration in this work. 
A definitive study of the $K^{-}pp$ quasi-bound state could be performed through fully exclusive 
formation reaction, such as the in-flight $\mathrm{^{3}He}(K^{-},N)$ reaction. This was performed at 
J-PARC~\cite{e15}. As a next step, we will develop the four-body Faddeev AGS equations to make a 
practical calculation of the cross section of kaon-induced strange-dibaryon production reaction. In 
the present study, we have calculated $K^{-}ppn$ and $K^{-}ppp$ quasi-bound state positions using 
the HSE method to find the separable expressions for the [3+1] and [2+2] subsystems. There is another 
separable expansion method for the [3+1] and [2+2] subsystems, this method is called the energy-dependent 
pole expansion (EDPE) method and the form factors have an energy dependence~\cite{sofia}. To study 
which one of these methods (HSE and EDPE) has a better convergence rate, one can perform a similar 
calculation using the EDPE method.

%%%%%%%%%%%%%%%%%%%%%%%%%%%%%%%%%%%%%%%%%%%%%%%%%%%%%%%%%% 
The authors thank A. Fix for helpful comments and discussions. One of the authors (S. Marri) is thankful 
to Prof. T. Yamazaki for his fruitful discussions. The authors gratefully acknowledge the Sheikh Bahaei 
National High Performance Computing Center (SBNHPCC) for providing computing facilities and time. SBNHPCC 
is supported by the scientific and technological department of presidential office and Isfahan University 
of Technology (IUT).
%%%%%%%%%%%%%%%%%%%%%%%%%%%%%%%%%%%%%%%%%%%%%%%%%%%%%%%%%%%
%%%%%%%%%%%%%%%%%%%%%%%%%%%%%%%%%%%%%%%%%%%%%%%%%%%%%%%%%%%%%%%%%%%%%%%%%%%%%%%%%%%%%%%%%%%%%%%%%%%


\bigskip
\begin{thebibliography}{}
\bibitem{akaishi} Y. Akaishi and T. Yamazaki, Phys. Rev. C {\bf 65}, 044005 (2002).
\bibitem{yamazaki1} T. Yamazaki and Y. Akaishi, Phys. Lett. B {\bf 535}, 70 (2002).
\bibitem{dote1} A. Dote, H. Horiuchi, Y. Akaishi and T. Yamazaki, Phys. Lett. B {\bf 590}, 51 (2004); Phys. Rev. C {\bf 70} 044313 (2004).
\bibitem{dote2} A. Dote, H. Horiuchi, Y. Akaishi and T. Yamazaki, Phys. Rev. C {\bf 70}, 044313 (2004).
\bibitem{shev1} N.V. Shevchenko, A. Gal and J. Mares, Phys. Rev. Lett. {\bf 98}, 082301 (2007).
\bibitem{shev2} N.V. Shevchenko, A. Gal, J. Mares, J. Revai, Phys. Rev. C {\bf 76}, 044004 (2007).
\bibitem{ikeda1} Y. Ikeda and T. Sato, Phys. Rev. C {\bf 76}, 035203 (2007).
\bibitem{ikeda2} Y. Ikeda and T. Sato, Phys. Rev. C{\bf 79}, 035201 (2009).
\bibitem{dote3} A. Dote, T. Hyodo and W. Weise, Nucl. Phys. A {\bf 804}, 197 (2008).
\bibitem{dote4} A. Dote, T. Hyodo and W. Weise, Phys. Rev. C{\bf 79}, 014003 (2009).
\bibitem{ikeda3} Y. Ikeda, H. Kamano and T. Sato, Prog. Theor. Phys. {\bf 124}, 533 (2010).
\bibitem{maeda} S. Maeda, Y. Akaishi and T. Yamazaki, Proc. Jpn. Acad., Ser. B {\bf 89} (2013).
\bibitem{yamazaki2} T. Yamazaki and Y. Akaishi, Proc. Jpn. Acad. Ser. {\bf B 83}, 144 (2007).  
\bibitem{yamazaki3} T. Yamazaki and Y. Akaishi, Phys. Rev. {\bf C 76}, 045201 (2007). 
\bibitem{HL} W.Heitler and F. London, Z. Phys. {\bf 44}, 455 (1927).
\bibitem{FINUDA} M. Agnello {\it et al.}, Phys. Rev. Lett. {\bf 94}, 212303 (2005).
\bibitem{Ramos} A. Ramos, V.K. Magas, E. Oset and H. Toki, Nucl. Phys. {\bf A 804}, 219 (2008).
\bibitem{e27} T. Nagae {\it et al.}, J-PARC E27 proposal; Y. Ichikawa {\it et al.,} Prog. Theor. Exp. Phys. 021D01 (2015).
\bibitem{yamazaki4} T. Yamazaki {\it et al.}, Phys. Rev. Lett. {\bf 104}, 132502 (2010).
\bibitem{Kienle} P. Kienle {\it et al.}, Eur. Phys. J. {\bf A 48}, 183 (2012).
\bibitem{HADES} L. Fabbietti {\it et al.}, HADES data.
\bibitem{fix} A. Fix and H. Arenhovel, Phys. Rev. C {\bf 66}, 024002 (2002).
\bibitem{alt} E.O. Alt, P. Grassberger and W. Sandhas, Nucl. Phys. B {\bf 2}, 167 (1967).
\bibitem{alt2} P. Grassberger and W. Sandhas, Nucl. Phys. B {\bf 2}, 181 (1967).
\bibitem{nadro} I. M. Nadrodetsky, Nucl. Phys. A {\bf 221}, 191 (1974).
\bibitem{sofia} S. A. Sofianos, N. J. McGurk, and H. Fiedeldey, Nucl. Phys. A {\bf 318}, 295 (1979).
\bibitem{shev4} N.V. Shevchenko, Nucl. Phys. A {\bf 890-891}, 50 (2012).
\bibitem{shev5} N.V. Shevchenko, Phys. Rev. C {\bf 85}, 034001 (2012).
\bibitem{para} H. Zankel, W. Plessas, J. Haidenbauer, Phys. Rev. C {\bf 28}, 538 (1983).
\bibitem{bazi} M. Bazzi {\it et al.}, Phys. Lett. B {\bf 704}, 133 (2011).
\bibitem{gal} N. Barnea, A. Gal and E. Z. Liverts, Phys. Lett. B {\bf 712}, 132 (2012).
\bibitem{roman} Roman Ya. Kezerashvili and Sh. M. Tsiklauri, EPJ Web of Conferences {\bf 81}, 02022 (2014).
\bibitem{e15} M. Iwasaki {\it et al.}, J-PARC E15 proposal.
\end{thebibliography}
\end{document}